\documentstyle[aps,version2,preprint]{revtex}    % preprint form
\begin{document}

\draft

\begin{title}
Pseudoparticle-operator description of an interacting bosonic gas
\end{title}

\author{A. H. Castro Neto $^{1}$, H. Q. Lin $^{1,*}$,
Y.-H. Chen $^{2,3}$, and J. M. P. Carmelo $^{1,2,**}$}
\begin{instit}
$^{1}$ Department of Physics, University of Illinois at Urbana-Champaign\\
1110 West Green Street, Urbana, Illinois 61801-3080
\end{instit}
\begin{instit}
$^{2}$ Instituto de Ciencia de Materiales, CSIC, Cantoblanco,\\
SP - 28049 Madrid, Spain
\end{instit}
\begin{instit}
$^{3}$ Joseph Henry Laboratories of Physics, Princeton University\\
Princeton, New Jersey 08544
\end{instit}
\receipt{June 1994}

\begin{abstract}
We write the Hamiltonian of the Bose gas with two-body
repulsive $\delta$-function potential in a pseudoparticle
operator basis which diagonalizes the problem via the Bethe
ansatz. In this operator basis the original bosonic
interactions are represented by zero-momentum forward-scattering
interactions between Landau-liquid pseudoparticles.
We find that this pseudoparticle operator algebra is
complete: {\it all} the Hamiltonian eigenstates are
generated by acting pseudoparticle operators on
the system vacuum. It is shown that one boson of
vanishing momentum and energy is a composite of a
one-pseudoparticle excitation and a collective
pseudoparticle excitation. These excitations have finite opposite
momenta and cannot be decomposed. Our formalism
enables us to calculate the various quantities which
characterize the static and dynamic behavior of the
system at low energies.
\end{abstract}
\renewcommand{\baselinestretch}{1.656}   % for preprint style only


\narrowtext
%%%%%%%%%%%%%%%%%%%%%%%%%%%%%%%%%%%%%%%%%%%%%%%%%%%%%%%%%%%%%%%%
\section{INTRODUCTION}

The interest on exact solutions of many-body problems has
increased in the last years due to the ``non-conventional" behavior
of new materials which cannot be described by the usual
perturbative approaches. The dynamics of
elementary excitations in these systems is constrained
to low-dimensional manifolds, one (for instance,
in quasi-one dimensional conducting polymers and
synthetic metals or edge states of the fractional quantum Hall
effect) or two (as in the cuprates) spatial
dimensions. It is known that the usual
perturbation theory does not give correct results since in low
dimensions (and specially in one dimension) the characteristics
of the elementary excitations, due to phase-space considerations,
change abruptly when the interaction between the particles is
turned on.

The lack of exact solutions in higher dimensions contributed
to the interest on one-dimensional integrable systems. These
have been used as a paradigm to understand what is called today
strongly interacting problems. The low dimensionality of these
systems implies that a weak interaction is dramatically enhanced,
which can be naively understood from the fact that particles
in one dimension cannot ``go around" each other (as they can in
higher dimensions).

On the one hand, exact solutions in one dimension (specifically, the
Bethe-ansatz [BA] solutions)
have been considered often too ``hermetic" to reveal the
full physical information contained in them. This misconception
follows from the fact that elementary excitations of models
solvable by the BA seem to have non-universal features.
For instance, at a first glance, the bosonic model
\cite{Liniger,Lieb,Berezin,Guire,Yang,Korepinrev}
(which will be treated in this paper),
with its fermionic excitation spectrum described by real
BA rapidities and the $SO(4)$ Hubbard model \cite{Lieb68,Korepin91}
which has the ``strings'' (described by complex, non-real,
rapidities) as its elementary eta spin and spin
lowest-weight-state (LWS) and highest-weight-state (HWS)
excitations \cite{Essler}, have many different features.

However, it was noticed recently \cite{Carmelo92,Carmelo93}
that the universality of the models solvable by
the BA becomes clearer when we look at
the sectors of parameter space of lowest symmetry $U(1)$.
In the case of Hamiltonians with a non-Abelian symmetry
these sectors correspond to finite values of the
generalized ``magnetic fields'' associated with the diagonal
generators of the non-Abelian algebras. The presence
of these ``magnetic fields'' lowers the symmetry
of the quantum problem to $[U(1)]^{\nu}$, where
$\nu$ is both the number of fields and of the associated
$U(1)$ Cartan sub-algebras. (At finite
eigenvalues of the diagonal generators the symmetry of
the problem is a global phase, $U(1)$, in each of the
$\nu$ channels of interaction [eta spin, spin, etc].) A
central point is that the expressions of many physical
quantities in the sectors of symmetry $[U(1)]^{\nu}$ provides
the corresponding values of these quantities in the sectors of
higher symmetry \cite{Carmelo92,Carmelo93}.
In addition, in these sectors the different
solvable BA problems have common universal properties,
all gapless low-energy branches of excitations being
of pseudoparticle-pseudohole type \cite{Carmelo93}.

One of the first applications of the BA
was the study of a continuum problem within this universality
class, i.e. a problem with a $U(1)$ symmetry. This was the
solution of the  many-body problem of bosons interacting by
a two-body $\delta$-function potential
\cite{Liniger,Lieb,Berezin,Guire,Yang,Korepinrev}.
As a field theory this is the repulsive quantum nonlinear
Schr\"odinger model \cite{Thacker}. This was also the first model
to which the quantum inverse-scattering method (QISM) was applied
\cite{Sklyanin,Faddeev,Izergin,Korepin}.

It was firstly noted by Girardeau \cite{Gira} that, although the
original problem is given in terms of bosons, the excitation
spectrum has a fermionic character when the interaction is
infinitely strong (the bosons acquire a hard core).
Afterwards, Lieb \cite{Lieb} pointed out
that the {\it elementary excitations} of this model were not
{\it quasiparticles} in the sense of the Landau theory of the
Fermi liquid \cite{Pines,Baym} because quasiparticles decay in time while
the elementary excitations, related to the excitation spectrum
and, therefore, to the exact diagonalization of the problem,
{\it never} decay. However, at that time, no {\it operational}
method was proposed to deal directly with the
elementary excitations in alternative to the original bosons
of the problem. (As we have pointed out, in the bosonic basis the problem
is nonperturbative.) The thermodynamics of the model was studied
by Yang and Yang and also revealed the fermionic character
of the elementary excitations \cite{Yang}. The asymptotic
behavior of the correlation functions was studied by several
authors \cite{Izergin,Korepin,Popov,Haldane,Bogo}.

In this paper we deal directly with the {\it elementary excitations}
which in the $U(1)$ sectors of the systems solvable
by the BA can be generated by applying on the particle
vacuum pseudoparticle fermionic operators
\cite{Carmelo92,Carmelo93,Carmelo91}. In the present
model the pseudoparticle operators generate a complete
orthonormal basis that spans the whole Hilbert space.
(This is in contrast to models with non-Abelian symmetries
where the pseudoparticle algebra refers to the low-energy Hilbert
space only \cite{Carmelo93}.) One of the motivations of our
study is the simplicity of the model which has a simple
$U(1)$ Abelian symmetry.

In the above operator basis the elementary excitations
are described by a Landau liquid of fermionic pseudoparticles.
By a Landau liquid we mean here the generalization of the Landau's
Fermi liquids \cite{Pines,Baym} introduced in Refs.
\cite{Carmelo92,Carmelo93,Carmelo91}. (These works refer mainly
to the one-dimensional Hubbard model.)
As in the case of the class of fermionic integrable quantum
liquids of Refs. \cite{Carmelo93}, the Hamiltonian,
charge current, and other operators of the present model
can be written in the pseudoparticle-operator basis. At
fixed number of bosons, $N$, {\it all} the excited states can
be generated by products of one-pair ``pseudoparticle-pseudohole''
operators acting on the $N$ interacting ground state (exactly as
found by Lieb \cite{Lieb}).

The study of the present quantum problem in the pseudoparticle
basis gives some new insights and provides a deeper understanding
of the previously known aspects of the physics of the model.
In particular, we will discuss the relationship between Lieb's
excitation spectrum and the pseudoparticle basis. This
reveals that the pseudoparticle description is the natural
representation for the quantum system. Moreover, we show that
the original bosons of the theory are formed by two
pseudoparticle excitations of opposite momentum.
As in the case of the electrons of the Hubbard
chain \cite{Carmelo94}, these excitations cannot
be decomposed and the pseudoparticles {\it cannot}
be removed or added to the many-particle system.
They exist only inside that system, their
configuration occupancies describing the Hamiltonian
eigenstates. However, at infinite bosonic interaction (Girardeau's
case \cite{Gira}) the pseudoparticles become free entities.

In the pseudoparticle basis and at fixed number of
bosons, $N$, the ground state has a non-interacting
form \cite{Carmelo93}. The normal ordered Hamiltonian has
in that basis an infinite number of terms,
each corresponding to a pseudoparticle forward-scattering
order, as we will see in Sec. II. While the low-energy
physics requires considering the first two terms
only, a full study of the physics of the model
at all energy scales requires the use of all
infinite terms of the Hamiltonian. Elsewhere
we will study another integrable model of simple $U(1)$
Abelian symmetry. This is the Calogero-Sutherland model
associated with quantum chaos in Ref. \cite{Shastry}.
The use of our operator basis reveals that the normal-ordered
pseudoparticle Hamiltonian has for this model
a finite number of terms. This further simplifies the study of its physics.

The various low-energy physical quantities can be derived as
in the Landau theory of Fermi liquids. In particular, we can obtain the
compressibility and the specific heat of the system and
confirm that the pseudoparticle transport mass is density
independent and that the charge conductivity spectrum reduces
to the Drude peak, as required by the translational invariance
of the model. In the pseudoparticle basis and {\it at all} energy
scales the present quantum problem has zero-momentum
forward scattering {\it only}. While at high energy
the physics is determined by multipseudoparticle
forward scattering of all orders, the low-energy
quantities are determined by two-pseudoparticle
scattering only. This reflects the perturbative
character of the pseudoparticle basis \cite{Carmelo93}.
We also discuss the properties of the system at the critical
point (scaling limit), including the study of the operator
representation for the Virasoro algebra \cite{Carmelo93}.

The paper is organized as follows: In Sec. II we present the
pseudoparticle basis and how it can be obtained directly from
the BA equations. The elementary excitations and
pseudoparticle description of the bosons are discussed
in Sec. III. In Sec. IV we study the static and dynamical
properties and consider the low-temperature thermodynamics.
The operator description of the Virasoro algebras is
introduced in Sec.V. In Sec. VI we study the charge
instabilities of the system. Finally, Sec.VII contains the
concluding remarks.

%%%%%%%%%%%%%%%%%%%%%%%%%%%%%%%%%%%%%%%%%%%%%%%%%%%%%%%%%%%%%%%%
\section{THE PSEUDOPARTICLE OPERATOR BASIS}

Let us consider a set of particles with bosonic statistics
which interact via Dirac's delta function. We describe their
dynamics by the following Hamiltonian \cite{Liniger,Lieb,Yang}

\begin{equation}
\hat{H} = - \sum_{j=1}^{N} {\partial^2\over {\partial x_j^2}}
+ 2c\sum_{l>j} \delta (x_j - x_l) + \mu N \, ,
\end{equation}
where $x_j$ is the position of the $j^{th}$ particle,
$c$ gives the strength of the interaction (here
we are interested in the repulsive case, $c>0$), and $\mu$ is the
chemical potential (we work in units such that $\hbar = 1$ and
the bare mass is $m=1/2$).

Originally, the diagonalization of this problem was carried out by
means of the BA \cite{Liniger,Lieb,Berezin,Guire,Yang}. Later, it was
shown by Thacker \cite{Thacker} that the two-body scattering
problem of the Hamiltonian $(1)$ can be {\it solved} in {\it perturbation
theory} by considering {\it all} infinite set of diagrams. Furthermore, all
the eigenstates are in the continuum since the interaction is
repulsive (no bound states) and is localized in space.
In other words, the spectrum of the non-interacting problem and the
interacting problem ``look'' the same, exactly as predicted by the
formal theory of scattering \cite{Merz}. However, the
similarity of the two problems is only apparent because turning
on the interaction changes the character of the quantum numbers
which label the Hamiltonian eigenstates. In contrast to the
non-interacting $c=0$ problem, the new quantum numbers of the
interacting system show global shifts in their values upon
changing the number of bosons $N$ by odd integers. This
implies that the above spectrum corresponds to new
fermionic objects, the pseudoparticles, and not to the original
bosons, as we discuss below.

It will be useful to consider the second quantized form of
the Hamiltonian $(1)$,

\begin{equation}
\hat{H} = \int dx \left( \frac{\partial \phi^{\dag}(x)}{\partial x} \,
\frac{\partial \phi(x)}{\partial x} + \mu \, \phi^{\dag}(x) \phi(x) +
c \, \phi^{\dag}(x) \phi^{\dag}(x) \phi(x) \phi(x) \right) \, ,
\end{equation}
where the operators $\phi$ and $\phi^{\dag}$ obey the usual bosonic
algebra. Direct inspection of the Hamiltonian $(2)$ reveals that it
is invariant under a global $U(1)$ transformation,

\begin{eqnarray}
\phi(x) \to e^{i \theta} \phi(x) \\
\phi^{\dag}(x) \to e^{-i \theta} \phi^{\dag}(x) \, ,
\nonumber
\end{eqnarray}
where $\theta$ is an arbitrary phase. This Hamiltonian can
be diagonalized ($\hat{H}\mid \Psi\rangle = E \mid \Psi \rangle$)
by the following wave function (the BA wave function)
\cite{Thacker}

\begin{eqnarray}
\mid \Psi(k_1,k_2,....,k_N)\rangle = \int
\left[\prod_{j=1}^{N} e^{i k_j x_j} dx_j\right]
\prod_{j<n\leq N} \left[1 - {ic\over {k_j-k_n}}
sgn (x_n-x_j) \right]
\nonumber \\
\phi^{\dag}(x_j)\mid V \rangle \, ,
\end{eqnarray}
where $\{k_j\}$ is a set of $N$ rapidity numbers (they are
canonically conjugated to $\{x_j\}$),
$\mid V \rangle$ is the vacuum of the problem,
$sgn(x)$ is $1$ for $x>0$ and $-1$ for $x<0$, and $\Delta(k)$ is the
two-particle ``phase shift'' which is obtained from the
the equation,

\begin{equation}
e^{i\Delta(k)} = \frac{k - ic}{k + ic} \, .
\end{equation}
The wave function $(4)$ is a superposition of plane
waves which is modulated by step functions. This is a basic
characteristic of the BA wave function since in one dimension
it is possible to order the particles in a line as
$0 \leq x_1 < x_2 <...< x_N \leq L$, where $L$ is the length of
the system. The spectrum of the problem, as in any problem in
quantum mechanics, is obtained from the boundary conditions.
As usual, we assume periodic boundary
conditions and from the wave function we find

\begin{equation}
e^{i k_j L} = \prod_{n \neq j} e^{i \Delta(k_n-k_j)} \, ,
\hspace{1 cm} j=1,....,N \, .
\end{equation}
Equation $(6)$ can be rewritten in the following form,

\begin{equation}
q_j = k_j +
{1\over L}\sum_{l=1}^{N} \Delta (k_j-k_l)
\hspace{1cm} j=1,.....,N \, ,
\end{equation}
where we choose

\begin{equation}
\Delta (k)= - 2 \, \tan^{-1}\left( \frac{k}{c}\right) \, ,
\end{equation}
and due to the boundary conditions

\begin{equation}
q_j = {2\pi\over L} n_j \, ,
\hspace{0.75cm} n_j=0,\pm 1,\pm 2 ... \, ,
\hspace{0.4cm} N\, odd\, ;
\hspace{0.4cm} n_j=\pm 1/2,\pm 3/2 ... \, ,
\hspace{0.4cm} N\, even\, ,
\end{equation}
are the {\it true} quantum numbers of the problem.

The wave function $(4)$ does not only describes
the ground state of the system but {\it all}
Hamiltonian eigenstates. Each eigenstate corresponds
to a different occupancy configuration of the
quantum numbers $q_j$, as we discuss below. This is in
contrast to Hamiltonians with non-Abelian symmetries where
the BA equations refer to the LWS or HWS of
the corresponding algebras only \cite{Korepin91,Essler,Carmelo93}.
In the present case each Hamiltonian eigenstate is characterized by a
distribution of $N$ occupied quantum numbers $\{q_j\}$ over the
infinite $q_j$ available values, as we discuss below.
We call the $N$ occupied values {\it pseudoparticles}.
In the present model the number of bosons $N$
{\it equals} the number of pseudoparticles. Eqs. $(6)$ and $(7)$
refer to the $N$ occupied values
only, whereas the available set of numbers $(9)$ corresponds
to all integers or half integers. For each set of $N$
rapidity numbers $\{k_j\}$ there will be one, and only one,
set of $N$ quantum numbers $\{q_j\}$. This one to
one relation is defined by Eq. $(7)$. This equation defines
the values of the rapidities

\begin{equation}
k_j = k(q_j) \, .
\end{equation}
(It also defines, implicitly, the values of the
rapidities corresponding to the unoccupied values
of $q_j$.) Moreover, the energy of an arbitrary
Hamiltonian eigenstate is given by (substitute $(4)$ in
$(1)$)

\begin{equation}
E = \sum_{j=1}^{N} [k_j]^2 + \mu N  \, .
\end{equation}
The total momentum is

\begin{equation}
K = \sum_{j=1}^{N} k_j \, ,
\end{equation}
where the sums run over the $N$ occupied $\{q_j\}$ values only.
While the energy $(11)$ involves the rapidity function
$(10)$, the total momentum $(12)$ can be expressed as a
sum of the pseudomomenta only. Summing both sides of $(7)$ over
the particle index $j$ leads to

\begin{equation}
K = \sum_{j=1}^{N} q_j \, .
\end{equation}
This is a general feature of many models solvable
by BA. It follows from the fact that the
two-particle ``phase shift'' is antisymmetric with
respect to interchange of momentum.

A very important property of the BA wave function is that
it vanishes if two occupied pseudomomenta $q_j$ are equal
\cite{Korepinrev,Carmelo93,Thacker}. Suppose, for simplicity,
that we have two particles in the problem only. In this case
$N=2$ and there are two occupied $\{q_j\}$ values, which
we call $q_1$ and $q_2$. From $(7)$ it is easy to
find that

\begin{equation}
q_1 - q_2 = k_1 - k_2 + 4 \,
\tan^{-1}\left(\frac{k_1 -k_2}{c}\right) \, ,
\end{equation}
and, therefore, if $q_1 = q_2$ we conclude from $(14)$ that
$k_1 = k_2$ is a unique solution. Thus, from direct inspection
of $(4)$ we find $\Psi = 0$. It clearly implies the
fermionic character of the excitation spectrum, although the
wave function $(4)$ is {\it symmetric} with respect to its
arguments as we would expect for a bosonic wave function,
i.e. its symmetry has nothing to do with the nature of
the spectrum. However, the bosonic character of the
wave function $(4)$ has effects on the instabilities
of the one-dimensional quantum liquid as, for example,
the absence of the fermionic Peierls instability
\cite{Campbell}.
Although the fermionic pseudoparticles lead to
the existence of a pseudo-Fermi surface, the Peierls
instability only shows up at infinite repulsion
(or zero density) where the pseudoparticles become
free fermionic entities, as we find in Sec. VI.

At fixed $N$ the ground state of the system is obtained
by ordering the occupied set of pseudomomenta $\{q_j\}$
symmetrically around zero \cite{Lieb,Nuno} from
$-\pi [N-1]/L$ to $\pi [N-1]/L$ \cite{Carmelo93,Nuno}. The boundary
conditions of Eq. $(9)$ assure that the two pseudo-Fermi points are
symmetric both for $N$ odd and $N$ even. This observation, summed to
the fact that the wave function vanishes for equal quantum numbers, is
enough to define the excitation spectrum of the system in terms of
a {\it pseudo-Fermi sea}, which is completely filled in the ground
state. That spectrum can be generated from operators which
produce excited states by ``creating" new $q_j$'s out of the sea or
``annihilating" $q_j$'s inside the sea. We associate with each
{\it pseudoparticle} (occupied quantum number $q_j$) of
{\it pseudomomentum} $q_j$ fermionic operators $b^{\dag}_{q_j}$
and $b_{q_j}$ which create and annihilate, respectively, these
pseudoparticles and obey the usual fermionic algebra
\cite{Carmelo93},

\begin{equation}
\{b_q^{\dagger},b_{q'}\} = \delta_{q,q'} \, ,
\end{equation}
and all other anticommutators vanish. This pseudoparticle
operator algebra generates all the Hamiltonian
eigenstates from the vacuum $|V\rangle$ of Eq. $(4)$,
as we find below.

Since these states can, alternatively,
be described in terms of the original bosonic operator
algebra, as in Eq. $(4)$, basic properties of quantum
mechanics imply that {\it there is} a canonical transformation
which connects the pseudoparticle and bosonic operators.
Its evaluation is a very involved problem, yet in
this paper we are able to describe the original
bosons in terms of pseudoparticles in some limits.

The above pseudoparticle-algebra construction is absolutely
consistent with the properties of the wave function and the
excitation spectrum given by the BA. This is because
the BA solution {\it refers} to that operator algebra
\cite{Carmelo93}. In the presence
of external fields the pseudoparticles are the
transport carriers of the system \cite{Carmelo92,Campbell}.

{}From now on we will be interested in the thermodynamical
limit only, where the numbers of particles, $N$, and the length of
the system, $L$, goes to infinity but the density, $n = N/L$, is
kept constant. In this limit we replace the sums by integrals
over the pseudomomentum (see the periodic boundary conditions
$(9)$). The total energy, the total momentum, and the total
number of particles are given by,

\begin{equation}
\frac{E}{L} = \int_{-\infty}^{+\infty} \frac{dq}{2\pi} \,
N(q) \, (k(q)^2 + \mu) \, ,
\end{equation}

\begin{equation}
\frac{K}{L} = \int_{-\infty}^{+\infty} \frac{dq}{2\pi} \, N(q) \, q \, ,
\end{equation}
and

\begin{equation}
\frac{N}{L} = n = \int_{-\infty}^{+\infty} \frac{dq}{2\pi} \, N(q) \, ,
\end{equation}
respectively, where $N(q)$ is the distribution function of
the pseudomomenta for a given state. In particular, for
the ground state we have the filled sea up to the pseudo-Fermi
momenta, $\pm q_F$, and, therefore, the occupation number
is given by

\begin{equation}
N^0(q) = \theta(q_F - \mid q \mid)
\end{equation}
where, in the thermodynamic limit,

\begin{equation}
q_F = \pi [n - {1\over L}] \approx \pi n \, .
\end{equation}
The BA equation $(7)$ is rewritten as

\begin{equation}
q = k(q) + \frac{1}{\pi} \int_{-\infty}^{+\infty} dq' N(q') \,
\tan^{-1}\left(\frac{k(q)-k(q')}{c}\right).
\end{equation}

Since the number of pseudoparticles equals the number
of bosons, $N$, we have constructed a scenario where we can act
with our pseudoparticle operators on states which are
generated from the vacuum $|V\rangle$ of Eq. $(4)$.
For a canonical ensemble with a fixed number of
bosons $N$ the corresponding ground state is obtained
by filling the pseudo-Fermi sea \cite{Carmelo93,Nuno}

\begin{equation}
|0;N\rangle = \prod_{|q|<q_F} b_q^{\dagger}|V\rangle \, ,
\end{equation}
where $q_F$ was defined in $(20)$. The Hilbert space
associated with each canonical ensemble is spanned by the set of
Hamiltonian eigenstates $|ex;N\rangle$ which can be constructed
from the ground state $(22)$ as follows

\begin{equation}
|ex;N\rangle = \prod_{i,j=1}^{N_{ph}}[b_{q_j}^{\dagger}b_{q_i}]
|0;N\rangle \, .
\end{equation}
This is a state with $N_{ph}$ excited pseudoparticles
(and pseudoholes).

It is also natural to define the number operator
$\hat{N}(q)$ which counts the number of pseudoparticles

\begin{equation}
\hat{N}(q) = b^{\dag}_q b_q \, .
\end{equation}
When acting on {\it any} Hamiltonian eigenstate
this operator gives

\begin{equation}
\hat{N}(q)|ex,N\rangle = N(q)|ex;N\rangle \, ,
\end{equation}
where the eigenvalue $N(q)$ is $1$ for occupied
pseudomomentum values and $0$ otherwise. In the
particular case of the ground state $(22)$,
$N(q)=N^0(q)$, where $N^0(q)$ is given in Eq. $(19)$.

Equations $(16)$, $(17)$, and $(18)$
can be seen as eigenvalue equations in the Hilbert space we have
just constructed. Although the following
pseudoparticle-operator representation
is similar to that of multicomponent integrable quantum
systems studied in Ref. \cite{Carmelo93}, the simplicity
of the present model makes it constructive and
interesting for clarifying the physics
to describe here that representation in some
detail.

For instance, the total momentum operator
and the total number operator are defined as \cite{Carmelo93},

\begin{equation}
\hat{K} = \sum_q q \hat{N}(q)  \, ,
\end{equation}
and

\begin{equation}
\hat{N} = \sum_q \hat{N}(q)  \, ,
\end{equation}
respectively. When acting on any state of the Hilbert
space (with fixed number of pseudoparticles) these
operators give

\begin{equation}
\hat{K} |ex;N \rangle = K |ex;N \rangle \, ,
\end{equation}

\begin{equation}
\hat{N} |ex;N \rangle = N |ex;N \rangle \, ,
\end{equation}
where $K$ and $N$ are given in $(17)$ and $(18)$, respectively.

Moreover, from $(21)$ it is clear that $k(q)$ is a functional of
$N(q)$ (that is, for each state labelled by the set of
occupied $q_j$ values, $\{q_j\}$, we generate
a different set $\{k_j\}$). Since all the states are
eigenstates of the number operator, the  above construction
implies that they are also eigenstates of the operator
$\hat{k}(q)$ defined by the eigenvalue equation

\begin{equation}
\hat{k}(q)|ex;N \rangle = k(q)|ex;N \rangle  \, ,
\end{equation}
and, in particular, for the ground state $(22)$ we have that

\begin{equation}
\hat{k}(q)|0;N \rangle = k_0(q)|0;N \rangle \, .
\end{equation}

We conclude from Eq. $(30)$ that the BA equation
$(20)$ can be seen as an {\it operator} identity, where the
numbers are exchanged by the respective operators (since it will
be valid for {\it any} of the Hamiltonian eigenstates
which constitute a complete orthonormal basis in
the Hilbert space). This relates the rapidity operator,
$\hat{k}$, and the pseudoparticle operators
defined above. However, this relationship is not trivial
and in its full form can only be obtained by the explicit
solution of $(21)$. The Hamiltonian can be
rewritten in the pseudoparticle operator basis as
(see (11))

\begin{equation}
\hat{H} = \sum_{q} \hat{N}(q) \left(\hat{k}(q)\right)^2 +
\mu \hat{N} \, .
\end{equation}
However, after expressing $\hat{k}$ in terms
of the pseudoparticles operator $(24)$ this expression is
extremely complex. We can simplify the problem if we construct
a normal-ordered expansion which gives us the operator
$\hat{k}$ in terms of the operators $(24)$, as in the case
of multicomponent integrable quantum liquids \cite{Carmelo93}.

The ground state $(22)$ is non interacting from the point
of view of the pseudoparticles. This is not true ``from the
point of view'' of the rapidities since as we change one
or a few of the occupied pseudomomenta $\{q_j\}$
there is a ``backflow" in ``rapidity space'' which can change
all the corresponding $\{k_j\}$ configuration
\cite{Carmelo92,Thacker}. However, from the point of view
of the pseudomomentum  $q_j$ there is no backflow since these
are the true quantum numbers which label the Hamiltonian
eigenstates. At fixed values of $N$ the quantum problem
is perturbative with respect to {\it pseudoparticle}
scattering processes but it is {\it not} perturbative
with respect to {\it particle} processes \cite{Carmelo93}
(scattering of bosons). Therefore, a natural expansion is to
create a small number (much smaller than the total number of
pseudoparticles $N$) of pseudoparticle-pseudohole pairs close to
the pseudo-Fermi surface. Any excited state can be
described in terms of the occupation number, $N(q)$, which we
rewrite as \cite{Carmelo92,Carmelo91}

\begin{equation}
N(q) = N^0(q) + \delta N(q) \, ,
\end{equation}
where $N^0(q)$ is the ground state occupation number, $(19)$, and
$\delta N(q)$ is the small deviation from the ground
state occupancy configuration.
Equation $(33)$ has a physically meaningful representation
in terms of our operators. We define the operator
$:\hat{N}(q):$, such that

\begin{equation}
:\hat{N}(q): = \hat{N}(q) - N^0(q) \, .
\end{equation}
This is {\it normal ordered} relatively to the ground state
($\langle 0;N\mid :\hat{N}(q): \mid 0;N \rangle = 0$).
By definition,

\begin{equation}
N^0(q) = \langle 0;N \mid \hat{N}(q) \mid 0;N \rangle.
\end{equation}
The deviation $\delta N(q)$ from the ground state to
an excited state $|ex; N \rangle$ is simply given by

\begin{equation}
\delta N(q) = \langle ex; N|:\hat{N}(q):|ex; N\rangle \, .
\end{equation}
This deviation represents a change
in the state of the system from the ground state
$|0;N\rangle$ $(22)$ to an excited state $|ex; N\rangle$
$(23)$.

We can expand the BA equation $(21)$ self-consistently in
terms of the deviation operator, $:\hat{N}(q):$.
The operator $\hat{k}$ will be a functional of the deviation
operator (see Eqs. $(21)$ and $(30)$). Therefore, these
two operators commute with each other. It is possible to
show, after some straightforward but lengthy algebra, that
the relation between these operators can be written as
\cite{Carmelo92,Carmelo93}

\begin{equation}
:\hat{k} (q): = k_0 (:\hat{{\cal Q}}(q):) - k_0 (q) \, ,
\end{equation}
where $k_0 (q)$ is the ground-state eigenvalue of Eq. $(31)$
and the operator $:\hat{{\cal Q}}(q):$ is given in terms of
the expansion of the deviation operator in the form,

\begin{equation}
:\hat{{\cal Q}}(q): = \sum_{n=1}^{\infty}
\hat{{\cal Q}}_n(q) \, .
\end{equation}
The operator $\hat{{\cal Q}}_n(q)$ corresponds
to the $n$th pseudoparticle scattering order.
Introducing $(37)$ and $(38)$ in the BA
equation and expanding order by order
provides the expression for all infinite
terms of the expansion $(38)$. For example,
the first-order terms reads

\begin{equation}
\hat{{\cal Q}}_1 (q) = \sum_{q'}
:\hat{N}(q'):\Phi (q,q') \, .
\end{equation}
Here $\Phi(q,q')$ are the shifts in the phase of the
pseudoparticle of pseudomomentum $q'$ due to the
zero-momentum forward-scattering collision with the pseudoparticle
of pseudomomentum $q$. This phase shift can
be obtained self-consistently by direct substitution of the
above expansion in the BA equation. They are
related to the usual BA phase shifts,
$\bar{\Phi }\left(k,k '\right)$, which are expressed in
terms of the ground-state rapidity numbers as

\begin{equation}
\Phi\left(q,q'\right) =
\bar{\Phi }\left(k_0 (q),k_0 (q')\right) \, .
\end{equation}
In order to achieve consistency between our expansion and the
BA equation we can show that the phase shifts must
obey the following integral equation

\begin{equation}
\bar{\Phi }\left(k,k '\right)
= -{1\over {\pi}}\tan^{-1}\left({k - k '\over c}\right) +
\int_{-Q}^{Q} dk '' R(k - k '')
\bar{\Phi }\left(k '',k '\right) \, ,
\end{equation}
where the kernel is given by

\begin{equation}
R(k) = {1\over {\pi c}}\left({1\over {1 +(k/c)^2}}\right) \, ,
\end{equation}
and

\begin{equation}
Q = \pm k_0(\pm q_F) \, .
\end{equation}

It is worth noticing that higher order terms in the
expansion $(38)$ are exclusively given in terms of the
phase shifts defined in $(40)-(41)$. The physical reason
for that is that higher-order scattering processes can be
decomposed into two-pseudoparticle scattering processes.
$(38)$ is an expansion in the pseudoparticle scattering order
\cite{Carmelo93}. The infinite expansion
$(38)$ is valid for the large class of integrable quantum
system of Ref. \cite{Carmelo93} which refers
to contact particle interactions. The normal-ordered
pseudoparticle Hamiltonian has an infinite
number of terms and is of the form

\begin{equation}
:\hat{H}: = \sum_{i=1}^{\infty} \hat{H}_i  \, ,
\end{equation}
where $\hat{H}_i$ is the Hamiltonian term of
$i$ th pseudoparticle scattering order. The
first-order term reads

\begin{equation}
\hat{H}_1 = \sum_q :\hat{N}(q):\epsilon (q) \, .
\end{equation}
Here $\epsilon (q)$ defines the pseudoparticle band

\begin{equation}
\epsilon (q) = \mu + [k_0(q)]^2 + 2\int_{-Q}^{Q}dk k
\bar{\Phi }\left(k ,k_0(q)\right) \, .
\end{equation}
In Fig. 1 the pseudoparticle band $(46)$ is plotted as a
function of the density for $c=1$. This band is a slightly distorted
version of the non-interaction bosonic spectrum
and it approaches the latter when $n \to \infty$,
which is the non-interacting bosonic limit.

The pseudoparticle group velocity is given by

\begin{equation}
v(q) = \frac{d\epsilon (q)}{d q},
\end{equation}
and the velocity at the pseudo-Fermi points is defined as

\begin{equation}
v_F=v(q_F) \, .
\end{equation}

Up to first order in $\delta N$ the pseudoparticles
are free-fermionic entities which have a spectrum given by $(46)$.
Moreover, in the present thermodynamic limit the interacting
terms are {\it irrelevant} for excitations of
small momentum and low energy.
This free behavior was also noted in previous studies of
the model \cite{Lieb,Thacker}.

We can proceed in our
calculation and go beyond the first order term
\cite{Carmelo92,Carmelo93}. This gives the
spectrum for excitations involving a small {\it but
finite} density of pseudoparticles, as we see in
Sec. IV. Again, substituting the expansions in
the BA equation and insuring consistency, we find the
next order term

\begin{equation}
\hat{H}_2 = \sum_{q,q'} :\hat{N} (q)::\hat{N} (q'):
{1\over 2}f\left(q,q'\right) \, ,
\end{equation}
where

\begin{eqnarray}
f(q,q') & = & 2\pi v(q)\Phi (q,q') +
2\pi v(q')\Phi (q',q) \nonumber \\
& + & 2\pi v\sum_{j=\pm 1}
\Phi (jq_F,q)\Phi (jq_F,q') \, .
\end{eqnarray}
Note the similarity between expression $(49)$ and the
usual Landau expansion of Fermi-liquid theory \cite{Baym}.
{}From $(46)$ and $(49)$, we can define the renormalized
dispersion relation,

\begin{equation}
\breve{\epsilon}(q) = \epsilon (q) +
{1\over 2\pi}\int_{-\infty}^{\infty}dq'
\delta N(q')f(q,q') \, .
\end{equation}
Naturally, the functions $f(q,q')$ are interpreted as the
$f$ functions of the theory and determine the two-pseudoparticle
interactions of the quantum liquid. The form of expression
$(51)$ is an universal characteristic for the systems solvable
by the BA in the sectors of parameter space of lowest $U(1)$
symmetry \cite{Carmelo92,Carmelo93}. The main difference
between these models is the form of the spectral parameters of
the BA equations. While in the usual Fermi liquids the
quasiparticles exist close to the Fermi surface only,
and describe approximations to Hamiltonian eigenstates,
in the present Landau liquid \cite{Carmelo92,Carmelo91}
pseudoparticle-pseudohole pairs are {\it true}
Hamiltonian eigenstates for all energy scales and
generate the whole spectrum of the
system. In integrable systems with non-Abelian algebras this
is true only for the Hilbert sub space spanned by the
LWS or HWS of these algebras
which are described by real rapidities \cite{Carmelo92,Carmelo93}.
(In the lowest-symmetry sectors and at low energy that sub
space coincides with the full Hilbert space.)

Let us define the $f^{+1}$ function (related to the
interaction of pseudoparticles at the same side of the
pseudo-Fermi sea) and $f^{-1}$ function (related to
the interaction at opposite sides
of the pseudo-Fermi sea), as

\begin{equation}
f^{+1} = f(\pm q_F,\pm q_F) \, , \hspace{2cm}
f^{-1} = f(\pm q_F,\mp q_F) \, .
\end{equation}
As in a Fermi liquid, the symmetric and antisymmetric combinations
of these functions define the Landau parameters

\begin{equation}
F^i =
{1\over{2\pi}}\sum_{j=\pm 1}(j)^i f^j
\, , \hspace{1cm} i=0,1 \, ,
\end{equation}
which play a relevant role in the low-energy physics.

The Landau parameters $(53)$ appear in the low-energy expressions
in the form of the ``renormalized'' velocities

\begin{equation}
v^i = v + F^i = v [\xi^i]^2
\, , \hspace{1cm} i=0,1 \, ,
\end{equation}
where $\xi^1$ is the dressed charge \cite{Korepin79}.
$\xi^0$ and $\xi^1$ can be written in terms of
two-pseudoparticle phase shifts as

\begin{equation}
\xi^i  = 1 + \Phi (q_F, q_F)
+ (-1)^i \Phi (q_F, -q_F)  \, ,
\hspace{1cm} i=0,1 \, .
\end{equation}

In agreement with the results of Refs. \cite{Korepin,Haldane,Bogo},
we find that $\xi^0$ is the inverse of the dressed charge

\begin{equation}
\xi^i  = 1/\xi^{1-i} \, , \hspace{1cm} i=0,1 \, ,
\end{equation}
and thus the following Luttinger-liquid relation
\cite{Haldane81} holds true

\begin{equation}
v_0 v_1 = (v)^2 \, .
\end{equation}
Equations $(53)-(57)$ reveal that there is consistency
between the Luttinger liquid of Haldane
\cite{Haldane81} and the Landau-liquid character of
the BA integrable systems of lowest symmetry $U(1)$.
The Luttinger-liquid parameters $(56)$ which define
the three velocities of Eq. $(57)$ arise here
from the Landau parameters of Eqs. $(53)$ and
$(54)$.

Equation $(57)$ can be used to express the velocities $(48)$
and $(54)$ in terms of the dressed charge or its
inverse as

\begin{equation}
v = {2\pi n \over {[\xi^1]^2}} =
2\pi n [\xi^0]^2 \, ,
\end{equation}
and

\begin{equation}
v_0 = {[v]^2\over {2\pi n}} \, ,\hspace{1cm}
v_1 = 2\pi n \, ,
\end{equation}
in agreement with the results of Refs.
\cite{Liniger,Popov,Haldane,Bogo,Popov72}.
The $f^{\pm 1}$ functions $(52)$ and Landau parameters $(53)$
are plotted as a function of the density in Figs. $(2)$,
whereas the velocities $(48),(54)$ and the dimensionless
parameters $(55)$ are plotted
in Figs. 3 and 4, respectively. At $c$ finite
and $n>0$ we have $v_1>v>v_0$. (The density
dependence of the functions plotted in Fig. 2 is
commented in Sec. III.)

In contrast to other integrable quantum liquids
\cite{Carmelo93}, the simplicity of the present model can be
seen in the fact that all observables depend on one quantity
only, namely, $\gamma = c/n$, as was pointed out in Ref.
\cite{Lieb}. Also, the $U(1)$ Abelian symmetry assures
that the pseudoparticle operator algebra generates
the whole Hilbert space from the vacuum of the
theory. This drastically simplifies the physical
interpretation of the information contained in that
operator basis which refers to the BA solution.
In Sec. III we combine the data presented in
this section with the study of the elementary excitations to
extract new physical information about the quantum problem.
Although the closed-form analytic form of the boson -
pseudoparticle operator transformation
remains an open question, we can obtain
an interesting physical picture concerning the description of
the bosons in terms of pseudoparticles.

%%%%%%%%%%%%%%%%%%%%%%%%%%%%%%%%%%%%%%%%%%%%%%%%%%%%%%%%%%%%%%%%
\section{BOSONS, PSEUDOPARTICLES, AND ELEMENTARY EXCITATIONS}

The results of Sec. II confirm that the elementary excitations
of the bosonic gas are fermionic and reveal that
such gas can be described by a pseudoparticle Landau liquid.
This has similar properties to the usual Fermi liquids with
an important conceptual change, as it was already pointed
out in Ref. \cite{Lieb}. The elementary
excitations do not refer to quasiparticles. In our
language this means that the pseudoparticles are
not simple ``dressed particles''. We find below that
one boson is a collective excitation involving {\it all}
pseudoparticles of the pseudo-Fermi sea. In addition,
the pseudoparticle occupancy configurations define
eigenstates of the quantum problem at all energy scales.

Otherwise, in what concerns two-pseudoparticle properties
we use the well known machinery \cite{Pines,Baym} of the
conventional Fermi liquids to treat the physics of the
pseudoparticles: in contrast to the one-particle problem,
we find below that the boson pairs {\it are not}
collective pseudoparticle excitations and involve
two pseudoparticles only. This is because the
collective pseudoparticle excitations of each
boson ``cancel'' in this case. In particular,
excitations which do not change the number $N$
of pseudoparticles and bosons can be treated
as in a Fermi liquid.

The simplest case of the latter excitations are
the one-pair pseudoparticles eigenstates.
Equation $(23)$ reveals that the
low-energy excitations of the system with a fixed number of
particles can be generated from the ground state $(22)$
by pseudoparticle-pseudohole processes close to the pseudo-Fermi
surface, as in a Fermi liquid. An one-pair excitation is written
as

\begin{equation}
|q,k\rangle = b_{q+k}^{\dagger}b_q|0;N\rangle \, ,
\hspace{2cm} 0<|q|<q_F \, , \hspace{1cm} |q+k|>q_F \, .
\end{equation}
In the thermodynamic
limit the spectrum of this excitation does not involve
the $f$ function term of $(44)$, which leads to corrections of
order $1/L$ only. The spectrum reads

\begin{equation}
\Delta E = \langle q,k|:\hat{H}:|q,k\rangle
= \epsilon (q+k) - \epsilon (q) \, .
\end{equation}
This spectrum is shown in Fig. 5 for various densities and $c=1$.
Its boundaries are the continuum spectra
I and II obtained by Lieb \cite{Lieb}. The spectra I and II
correspond to removing and adding one boson from the
system, respectively. The reason for this seems straightforward:
since the number of pseudoparticles equals that of the bosons,
if we remove or add one boson it is equivalent to diminish
or increase the number of pseudoparticles by one and this
creates an extra pseudohole or pseudoparticle relatively to
the ground state $(22)$, respectively. However, there is a
{\it fundamental}
difference between the excitations which conserve the number
of pseudoparticles and bosons, $N$, as the one-pair
states $(60)$ and the general excitations $(23)$, and these
which change $N$ by one (or any odd number).

The present picture has some basic similarities with
the electrons of the Hubbard chain studied
in Ref. \cite{Carmelo94}. One boson of vanishing
momentum and energy includes two pseudoparticle excitations
which cannot be decomposed: (a) one
pseudoparticle of momentum $\pm q_F$; and (b) a
collective excitation of {\it all} the remaining
$N-1$ pseudoparticles, each contributing
with a small fraction $\mp \pi/L$ to the
momentum of the boson. Although in the
present thermodynamic limit each fraction
$\mp \pi/L$ is vanishing small, if we multiply
by the number of $N-1$ pseudoparticles of
the pseudo-Fermi sea this gives $\mp \pi [n - {1\over L}]$
which, following Eq. $(20)$, gives {\it precisely}
$\mp q_F$. The fact that one boson decays in
one pseudoparticle and a collective excitation
involving the remaining pseudoparticles justifies
the non-perturbative character of the bosonic basis
\cite{Carmelo93}.

Although the energy spectra I and II involve only the band
$\epsilon (q)$ of the corresponding ``missing'' or added
``pseudoparticle'', this type of state {\it is not}
an one-pseudoparticle excitation and affects {\it all}
pseudoparticles of the pseudo-Fermi sea. Adding or removing
one boson includes both adding and removing one
pseudoparticle and the above collective pseudoparticle
excitation and these two pseudoparticle excitations
{\it cannot} be decomposed. Therefore,
the pseudoparticles are not asymptotic states of
the many-body system. Although their occupancy configurations
describe the Hamiltonian eigenstates and they are the transport
carriers and couple to external fields
\cite{Carmelo92,Campbell}, they are ``confined'' to
the many-particle system. The only entities
which we can add or remove are the bosons. These
are the true asymptotic states of the problem. Since the number
of bosons and pseudoparticles are equal,
removing or adding one boson diminishes or increases
the number of pseudoparticles by one. However, this
``one-boson excitation'' has in the pseudoparticle basis
a collective character.

On the other hand, the pseudoparticle-pseudohole state
$(60)$ is a one-pair pseudoparticle excitation which involves the
transfer from the pseudo-Fermi sea to the unoccupied
pseudo-Brillouin zone of one pseudoparticle only. This
excitation does not affect the remaining pseudoparticles of
the pseudo-Fermi sea because it does not change $N$.
Furthermore, removing two (or
an even number $\cal{N}$) of bosons can be seen, in the pseudoparticle
basis, as a two-pseudoparticle (or $\cal{N}$-pseudoparticle)
excitation. This is because in this case there is no global
pseudomomentum shifting, as confirmed by Eq. $(9)$.

The spectrum I (or II) of Ref. \cite{Lieb} is degenerated with
an excitation where we transfer one pseudoparticle from a
pseudomomentum $q<q_F$ to $q_F^{+}$ (from $q=q_F$ to
$q>q_F$). This is the lower (or upper) boundary of Fig. 5.
Following the boundary
conditions of Eq. $(9)$, the $N$-particle and
$N\pm 1$-particle ground states $(22)$ have both
zero momentum. This is because the pseudomomentum of
the ``removed'' or ``added'' pseudoparticle is compensated
by the global pseudomomentum shift $\pm \pi/L$,
as we have discussed above.
Since this collective pseudoparticle state involves
$N\pm 1$ pseudoparticles, it has an excitation momentum
of $\pm q_F$, as confirmed by Eq. $(20)$. Therefore,
the excitations I and II can be decoupled into two
excitations: (i) a change from the $N$ ground state
to the $N\pm 1$ ground state. In the thermodynamic
limit this excitation has both vanishing energy
and momentum; (ii) the above pseudoparticle-pseudohole
excitation involving transfer of one pseudoparticle to or
from the pseudo-Fermi surface of the $N\pm 1$ system.

If we remove two bosons of vanishing
energy and momentum from the many-body system,
the two collective pseudomomentum shifts have opposite
momenta and cancel. Therefore, this can be seen as
a two-pseudoparticle excitation.
Since both the individual bosons and the
pair of bosons have vanishing momentum,
the pair of bosons is constituted by
two pseudoparticles of pseudomomenta
$q_{F}$ and $-q_{F}$. Therefore, we expect
the pseudoparticle Hamiltonian to
favour attraction between such pair and
repulsion between two pseudoparticles with the same
pseudomomentum $\pm q_F$.

This is confirmed by the signs of the two-pseudoparticle
$f^{\pm 1}$ functions $(52)$ plotted in Fig. 2. When
$\gamma \to 0$ the system is non interacting
from the point of view of the bosons, whereas when $\gamma \to \infty$
the bosons are strongly interacting and we obtain the case
studied by Girardeau \cite{Gira}. By solving numerically the
BA equations for a fixed $c$ and varying the
density $n$ we obtain the behavior for the
$f^{\pm 1}$ functions $(52)$. Figure 2 confirms
that when the density increases (or the interaction between the
bosons decreases) the two-pseudoparticle repulsion increases
in one channel ($f^1$) and the two-pseudoparticle attraction
increases in the other channel ($f^{-1}$). Moreover,
the pseudoparticle attraction always dominates the dynamics
of the system at large densities. The interpretation
of this behavior is straightforward from the point
of view of pseudoparticles: when the bosons interact strongly
(Girardeau's case, $n \to 0$) the pseudoparticles tend to
become free and their excitation spectrum is the one of
free fermions. For finite values of the bosonic interaction
the two pseudoparticles with opposite momentum
which constitute a bosonic pair attract each other.
This gives rise to bound states (analogous to Cooper pairs of the
BCS theory of superconductivity). This attraction becomes
more pronounced when the bosons interact weakly ($n \to
\infty$). This confirms that the above bosons pairs
of vanishing energy and momentum are nothing but pairs of
pseudoparticles of opposite momentum.

Moreover, the attraction between the two pseudoparticles
of pseudomomenta $q_{F}$ and $-q_{F}$ and the repulsion
between the pairs of pseudoparticles with same
pseudomomentum $\pm q_{F}$ is associated with
the two collective pseudoparticle excitations of
the individual bosons which constitute the pairs removed
or added to the system. These collective pseudoparticle
excitations are due to the shift between
the two boundary conditions of $(9)$. Note
that {\it only} when $\gamma$ is infinite the
function $(8)$ vanishes and Eq. $(6)$ does not lead,
necessarily, to the two boundary conditions of $(9)$:
in this case the numbers $n_j$ can be integers both
for $N$ even and odd, as in a non-interacting system.
However, any finite value of $\gamma$ implies the
occurrence of the two boundary conditions $(9)$
which are controlled by the parity of $N$.
Therefore, when $1/\gamma >0$ the bosons always include
the collective pseudoparticle excitation. At finite $c$
the equality $1/\gamma =0$ implies zero density, $n=0$, and
there are no bosons or pseudoparticles left.
When $1/c=0$ the infinite interaction between the bosons
allows the pseudoparticles to be completely decoupled
and they form a free-fermionic gas, as it was
pointed out by Girardeau.

Our study justifies the crossover between the
elementary excitations of this system: non-interacting bosons
in one limit, and non-interacting fermionic pseudoparticles in
the opposite limit.

%%%%%%%%%%%%%%%%%%%%%%%%%%%%%%%%%%%%%%%%%%%%%%%%%%%%%%%%%%%%%%%%
\section{STATIC AND TRANSPORT QUANTITIES}

Apart from the appealing physical interpretation of the
bosons in terms of pseudoparticle excitations, which by itself
is enough reason for the introduction of the formalism
presented here, we can use the same procedures as in Landau's
Fermi liquid theory to calculate measurable quantities related
to the response of the system to external fields.

While the study of the dynamic correlation functions
at all energy scales requires considering all
pseudoparticle scattering orders of the Hamiltonian
$(44)$, the perturbative character of the pseudoparticle
basis determines that the low-energy physics is
solely controlled by the first two terms of
that Hamiltonian, as we have discussed in
previous sections.

In what concerns excitations conserving the pseudoparticle
number $N$ the present pseudoparticle liquid is very similar
to the usual Fermi liquids of quasiparticles.
As a simple example of the analogy with Fermi liquids we
calculate the compressibility of the interacting bosonic gas. The
compressibility is obtained as the change in the chemical
potential due to the change in the number of bosons.
This refers to the energy changes of the ground states
with $N-1$ and $N+1$ bosons. Since $N-1$ and $N+1$ have the
same parity, Eq. $(9)$ assures that the corresponding
elementary excitation is of both two-pseudoparticle
and two-boson character, and does not involve
pseudoparticle collective excitations of the type
discussed in Sec. III. Therefore, this excitation is equivalent to
the change in the pseudo-Fermi momentum $q_F$ by a small
quantity $\delta q_F=\pi\delta n$ without changing the integer
or half-integer character of the numbers of Eq. $(9)$.

In terms of the distribution functions $(33)$ and $(36)$
we can write,

\begin{equation}
N(q)= \Theta\left(q_F + \delta q_F - |q|\right) \, ,
\end{equation}
and, up to first order in $\delta q_F$ we obtain

\begin{equation}
\delta N(q) = \delta (q_F - |q|)
\delta q_F \, .
\end{equation}
Following a calculation analogous to the one performed
in Refs. \cite{Carmelo91}, we easily obtain the change of
the chemical potential with the density as

\begin{equation}
{\partial \mu(n)\over {\partial n}} = - v_0\pi \, ,
\end{equation}
where $v_0$ is defined in Eq. $(54)$. The
compressibility is easily obtained and reads

\begin{equation}
\chi  = - {1\over {n^2}} {1 \over
{\partial \mu (n)/\partial n}}
= {1\over {\pi n^2}} {1 \over {v_0}} \, .
\end{equation}
Using the asymptotic values for the parameter $v_0$ given in
Table 1, we find that at fixed value of the interaction
the compressibility diverges as $n^{-3}$ in the low density limit
when $n \to 0$ (strongly interaction of bosons). It goes to zero
as $n^{-2}$ in the large density limit $n \to \infty$ (weakly
interaction of bosons). This behavior is clearly expected
since there is a crossover from fermionic to bosonic
excitations. The inverse of the compressibility is plotted
in Fig. 6 as a function of the density $n$.

Also the low-temperature thermodynamics can be studied
as in a Fermi liquid. This leads to the same results as
Yang in Ref. \cite{Yang}. The entropy is given in terms of
the distribution function $N(q)$ as

\begin{equation}
S = - {L\over {\pi }}\int_{-\infty}^{\infty}dq
\{N(q)\ln [N(q)] + (1-N(q))\ln [1-N(q)]\} \, .
\end{equation}
It follows that the distribution function at finite small temperature
$T$ is given by

\begin{equation}
N(q) = {1\over {1+e^{\frac{\epsilon (q)}{k_BT}}}} \, .
\end{equation}
(Note that $\epsilon (q_F)=0$.) From $(66)$ and $(67)$ it
is straightforward to obtain the low-temperature specific heat,
which reads

\begin{equation}
c_V/L = \left({k_B^2\pi \over 3}\right) {1\over {v}} T
= \left({k_B^2 \over {3n }}\right) m^* T
\, ,
\end{equation}
where we have defined the effective mass to be

\begin{equation}
m^* = {q_F \over {v}} \, ,
\end{equation}
and $v$ is given in Eq. $(48)$. At fixed
temperatures the specific heat diverges as
$n^{-1}$ at low densities and goes to zero as $n^{-1/2}$ at
high densities (see Table 1).

These results are consistent with the
non-interacting limit. At finite $c$ the non-interacting bosonic
system corresponds to the infinite-density limit. Since the
total energy is always finite, the chemical potential must vanish
in that limit. On the other hand, the dispersion relation for the
free bosons with finite chemical potential is given by

\begin{equation}
E_k = \sqrt{k^2 + \mu} \, ,
\end{equation}
which reproduces exactly the result $(68)$ for zero chemical
potential with mass $m = 1/2$, as expected.

Following the same route as in the calculation of the static
properties we can also use the transport equations of the usual
Fermi liquids to study the transport properties of the present
Landau liquid. In particular, we are interested here in the
conductivity and corresponding transport mass.

The expression of the normal-ordered Hamiltonian $(44)$,
whose first two terms are given in Eqs. $(45)$ and
$(49)$, is a sum of integrals over products of the
pseudomomentum distribution operator $(34)$. We
find that the charge-current operator is also
of that form. It contains zero-momentum
pseudoparticle forward-scattering terms
only and commutes with that Hamiltonian.
It follows that the conductivity spectrum has no
incoherent part and is constituted by the Drude $\delta$ peak
only. This is consistent with the translational invariance of
the system which implies that the conductivity sum rule
\cite{Baeri} involves the bare mass $m=1/2$ and
reads

\begin{equation}
\int_{0}^{\infty} d\omega \, Re \sigma (\omega ) =
{\pi n\over {m}} = 2\pi n \, .
\end{equation}
Here we have used units of charge equal to $1$.
Since, following the translational invariance of the
system, the sum rule $(71)$ has to be exhausted by the
Drude peak, the conductivity spectrum reads

\begin{equation}
Re \, \sigma(\omega) = 2 \pi D \delta(\omega) \, ,
\end{equation}
where, combining Eqs. $(71)$ and $(72)$, we have
\cite{Kohn}

\begin{equation}
D = n = \frac{q_F}{2 \pi m} \, ,
\end{equation}
and $m=1/2$.

To confirm that the pseudoparticles are
the transport carriers, we rederive the spectrum
$(72)$ via kinetic equations, as in Ref. \cite{Carmelo92}
for the case of the Hubbard chain. This confirms
the expected relation between the velocity
$v_1$ $(54)$ and the stiffness. As is shown in Ref.
\cite{Carmelo92}, that parameter defines the elementary
pseudoparticle current. The same method as in that reference
leads to the conductivity spectrum $(72)$ with

\begin{equation}
2\pi D = v^1 \, .
\end{equation}
Since, following Eq. $(59)$, $v^1 = 2\pi n$, we thus conclude
that the pseudoparticle
transport mass \cite{Carmelo92} {\it coincides} with the bare
bosonic mass $m=1/2$, as required by translation invariance.

%%%%%%%%%%%%%%%%%%%%%%%%%%%%%%%%%%%%%%%%%%%%%%%%%%%%%%%%%%%%%%%
\section{Conformal invariance}

The properties of the system at low energies can be investigated
from the point of view of conformal-field theory. Here
we follow the operator analysis of Ref. \cite{Carmelo93}.
The suitable low-energy critical-point Hamiltonian is
constructed in the pseudoparticle basis. Using the same
procedures as in Ref. \cite{Carmelo93}, we linearize the bands
of $(45)$ and replace the $f$ functions of $(49)$ by the values
$(52)$ to obtain

\begin{eqnarray}
:\hat{H}: & = & \sum_{\kappa,\iota=\pm 1}
\iota \kappa v :\hat{N}_{\iota}
(\kappa): + \frac{1}{2 L} \sum_{\kappa,\kappa'} \sum_{\iota=\pm 1}
[f^{1} :\hat{N}_{\iota}(\kappa)::\hat{N}_{\iota}(\kappa'):
\nonumber\\
& + & f^{-1} :\hat{N}_{\iota}(\kappa)::\hat{N}_{-\iota}(\kappa'):]
\, ,
\end{eqnarray}
where the pseudo-wavevectors are written relative to the pseudo-Fermi
momentum as $\kappa =q-\iota q_F$. Here $\iota =sgn (q)1$
defines the $\iota =1$ right and $\iota=-1$ left movers.
The operator $b_{\kappa ,+1}$ (or $b_{\kappa ,-1}$)
refers to the annihilation of a pseudoparticle moving to the
right (or left) with pseudomomentum $\kappa$ relative to the
corresponding pseudo-Fermi point $q_F$ (or -$q_F$).
In $(75)$ the $f^{\pm}$ functions $(52)$ can be expressed in terms
of the velocity $(48)$ and dimensionless parameters $(56)$
as follows

\begin{equation}
f^{\iota} = 2 \pi v \left(-\delta_{\iota,1}
+ \frac{1}{2} \left[(\xi^0)^2+\iota (\xi^1)^2\right] \right)
\, .
\end{equation}
(These functions are plotted in Fig. 2.) The only scale in
the quantum problem $(75)$ is the pseudo-Fermi
velocity, $v$, which, following $(76)$, is an overall
multiplication factor in the Hamiltonian $(75)$ (all other
parameters are dimensionless).

The Hamiltonian $(75)$ can be rewritten
as a pseudoparticle Luttinger liquid with right and
left potentials given by $\delta_{\kappa,0}f^{\pm1}$
\cite{Carmelo94}. Haldane \cite{Haldane81} has shown that the
low-energy spectrum of some of the systems solvable by BA
can be mapped in that of the Luttinger model. Here we have
shown that the pseudoparticle basis provides an explicit
{\it operator} construction for that fact.

At the critical point the HWS of the Virasoro algebra
are generated by the following type of states:
(A) Hamiltonian eigenstates which have a small density of
pseudoparticles or pseudoholes added to the ground state,

\begin{eqnarray}
|(A)\rangle = \prod_{\iota} \prod_{\kappa=0}^{\iota \delta q_F}
b^{\dag}_{\kappa,\iota} |0;N\rangle
\hspace{1cm} \, \delta q_F > 0 \, ,
\nonumber
\\
|(A)\rangle = \prod_{iota} \prod_{\kappa=0}^{\iota \delta q_F}
b_{\kappa,\iota} |0;N\rangle
\hspace{1cm} \, \delta q_F <0 \, ,
\end{eqnarray}
where $\delta q_F = \pi \delta n$ is the change in the pseudo-Fermi
momentum due to the corresponding change in the bosonic density;
and (B) Hamiltonian eigenstates associated
with a density of pseudoholes in one side of the pseudo-Fermi surface
and the same density of pseudoparticles in the opposite side of
that surface,

\begin{equation}
|(B)\rangle = \prod_{\kappa=0}^{\tilde{\delta} q_F}
b^{\dag}_{\kappa,\iota}
b_{\kappa,-\iota} |0\rangle
\hspace{1cm} \, \iota q_F > 0 \, .
\end{equation}
Here we have defined $\tilde{\delta} q_F = \frac{2 \pi {\cal D}}{L}$,
where $2{\cal D}=\delta N_{\iota =1}-\delta N_{\iota =-1}$
is related to the number of pseudoparticles transferred
across the pseudo-Fermi sea. $N_{\iota}$ is the
number of right ($\iota =1$) and left
($\iota =-1$) carriers. The Hamiltonian eigenstates
(B) have large momentum $K={\cal D}2q_F$.

The exclusive zero-momentum forward-scattering character
of the Hamiltonians $(44)$ and $(75)$ implies separated
conservation laws for the numbers $N_{1}$ and $N_{-1}$.
Both the excitations (A) and (B) change these numbers. Note,
however, that excitations (A) conserve the number
$N_{1}-N_{-1}$, whereas excitations (B) conserve
the number $N=N_{1}+N_{-1}$.

The tower of excitations, which conserve both
numbers $N_{\iota}$, can also be written in terms of the
pseudoparticles operators. We denote these excitations by
states (C) which involve a density of
pseudoparticle-pseudohole pairs around the same pseudo-Fermi
point. These are small-momentum and low-energy
excitations given by

\begin{equation}
|(C)\rangle = {\cal L}^{\iota}_{-N_{ph}^{\iota}} |0;N\rangle \, ,
\end{equation}
where ($j<0$)

\begin{equation}
{\cal L}^{\iota}_{j} = \prod_{\kappa_{p,\iota}} \prod_{\kappa_{h,\iota}}
b^{\dag}_{\kappa_{p,\iota},\iota}
b_{\kappa_{h,\iota},\iota} \, ,
\end{equation}
are the generators of the Virasoro algebra. In Eq. $(79)$

\begin{equation}
N_{ph}^{\iota} =  \iota \frac{L}{2 \pi} \left( \sum_{p}
\kappa_{p,\iota} - \sum_{h} \kappa_{h,\iota}\right) \, ,
\end{equation}
is the total number of pseudoparticle-pseudohole pairs
on the side $\iota$ of the pseudo-Fermi sea. This
quantity is given in terms of the net momentum involved
by the creation of one pseudoparticle with momentum
$\kappa_{p,\iota}$ and the creation of a pseudohole with
momentum $\kappa_{h,\iota}$ across the pseudo-Fermi
surface. If we consider states with a fixed number of
particles and small momentum, only the states (C)
contribute.

It is easy to show that the Virasoro generator of zero order,
which is related to spatial and temporal translations,
is given by (we set $v=1$)

\begin{eqnarray}
{\cal L}^{\iota}_{0} & = & \iota \sum_{\kappa} \kappa
:\hat{N}_{\iota}(\kappa):
+ \frac{1}{L} \sum_{\iota'}\sum_{\kappa,\kappa'}
[({f^{1}\over {2\pi}} + {\iota\iota'\over 4})
:\hat{N}_{\iota'}(\kappa): :\hat{N}_{\iota'}(\kappa'):
\nonumber\\
& + & ({f^{-1}\over {2\pi}} + {\iota\iota'\over 4})
:\hat{N}_{\iota'}(\kappa): :\hat{N}_{-\iota'}(\kappa'):]
\, .
\end{eqnarray}

The excitation energy $\Delta E$ and momentum $K$ corresponding
to Hamiltonian eigenstates involving all three types of
processes as in $|(A)\rangle$, $|(B)\rangle$, and
$|(C)\rangle$ is written as

\begin{equation}
\Delta E = \langle : \hat{H} : \rangle =
{2\pi\over L}\sum_{\iota} v\left(
h^{\iota} + N^{\iota}_{ph} \right) \, ,
\hspace{1cm} K = {2\pi\over L}\sum_{\iota}\iota\left(
h^{\iota} + N^{\iota}_{ph} \right) +
{\cal D}2q_F \, ,
\end{equation}
where $h^{\iota}$ are the dimensions of the fields which
are given by

\begin{equation}
h^{\iota} = \frac{1}{2} \left( \xi^1 {\cal D} +
\iota\xi^0 \frac{\delta N}{2}\right)^2 \, .
\end{equation}

Our pseudoparticle operator basis has allowed the study of the
Virasoro algebra. Both the HWS and the towers states (C) are
generated by acting the pseudoparticle operators on the ground
state. This has the non-interacting form given in
Eq. $(22)$. We have written the generators of the
algebra and other operators. Only the present
pseudoparticle basis allows a simple operator representation
for the generators of Virasoro algebras of BA
integrable models \cite{Carmelo93}.

The Virasoro algebra can also be used to evaluate the
asymptotic of the correlation functions. We omit here that study.
These functions have been obtained by other authors
\cite{Izergin,Korepin,Popov,Haldane,Bogo}.

%%%%%%%%%%%%%%%%%%%%%%%%%%%%%%%%%%%%%%%%%%%%%%%%%%%%%%%%%%%%%%%%
\section{CHARGE INSTABILITIES}

The formalism we have presented in the previous sections
also allows us finding and classifying the possible divergence
in the response functions \cite{Campbell}. This
provides the instabilities of the many-boson system.
For instance, the real part of the charge-charge response
function at zero frequency can be written in terms of the
dynamic-form factor $S(k,\omega)$ \cite{Pines} as

\begin{equation}
Re\, \chi(k,0) = -2 \, \int_0^{\infty} d\omega
\frac{S(k,\omega)}{\omega}\, .
\end{equation}
Using the methods of Ref. \cite{Campbell} we can show that in the limit
of small frequency and $k= {\cal D}2q_F$
the dynamical form factor is given by

\begin{equation}
S({\cal D}2q_F,\omega) = S_0({\cal D}2q_F) \omega^{\zeta(
{\cal D}2q_F)} + h.o.t. \, ,
\end{equation}
where the higher-order terms vanish as $\omega\rightarrow 0$
and $k={\cal D}2q_F$ are the only values of momenta at
which $S(k,\omega)$ can be non-vanishing at small $\omega$
\cite{Campbell}. In Eq. $(86)$ $\zeta$ are the exponents which classify the
divergence of the function $(85)$. If $\zeta>0$ there is no
divergence in $(85)$ and $S({\cal D}2q_F,\omega)$ vanishes at
$\omega =0$, if $\zeta=0$ there is a logarithmic divergence, and
if $\zeta<0$ there is a power-law divergence. The exponent
$\zeta$ is related to the dressed charge $\xi^1$ $(55)$ by

\begin{equation}
\zeta({\cal D}2q_F) = 2 [(\xi^1 {\cal D})^2 -1]\, .
\end{equation}
{}From the use of the solution of the BA equation for the
phase shifts $(41)$ we find that $\zeta$
is always greater than zero except for ${\cal D} = \pm 1$
and (a) for vanishing density at finite $c$; and (b) for
infinite bosonic repulsion at finite $n$. In these
limits $(87)$ gives $\zeta(2q_F) = 0$. Therefore, in these particular
limits and at the momenta $k=\pm 2q_F$ we have a logarithmic
singularity in the response function $(85)$. This singularity
is expected for a fermionic gas. It is known as the Peierls instability
\cite{Campbell}. Again, the free-fermionic character of a gas of
bosons with hard cores appears in the response function and,
therefore, it can be measured experimentally. For
all other values of the interaction and density the
bosonic character of the Hamiltonian eigenstates $(4)$
``kills'' the Peierls instability.

%%%%%%%%%%%%%%%%%%%%%%%%%%%%%%%%%%%%%%%%%%%%%%%%%%%%%%%%%%%%%%%%
\section{CONCLUDING REMARKS}

In this article we have shown that a gas of bosons interacting
repulsively via hard core potentials can be seen as a liquid of
interacting fermionic pseudoparticles. The bosons are at
vanishing energy and momentum an object constituted by
two excitations which {\it cannot} be decomposed: one
pseudoparticle of pseudomomentum $\pm q_F$ and a collective
pseudoparticle excitation of momentum $\mp q_F$. In addition,
one pair of bosons of vanishing momentum and energy
is constituted by a pair of pseudoparticles with opposite
momentum $q_F$ and $-q_F$, which form a bound state in the
non-interacting bosonic limit. At infinite bosonic interaction
the pairs break due to the scattering between the bosons.
This gives rise to a gas of free fermionic pseudoparticles.
In spite of the fermionic character of the excitations,
the bosonic nature of the Hamiltonian eigenstates implies
that only in this limit the system becomes really fermionic
and develops a Peierls instability at the momenta
$\pm 2q_F$.

Although some of the features of the model were already studied
by many authors, we have presented a new operator representation
for the problem. This uses a consistent framework
valid in the $U(1)$ sectors of parameter space of all
systems solvable by the BA. (In the present model
this refers to all the parameter space.)
Analogously to what happens in the sectors of lowest
symmetry of other models as the Hubbard
chain, the elementary excitations are described by a liquid of
pseudoparticles with zero-momentum forward-scattering
interactions only.

Our operator pseudoparticle algebra generates {\it all}
Hamiltonian eigenstates from the bosonic vacuum.
The pseudoparticle algebra
refers to the BA solution whose
equation was solved to provide a perturbative
normal-ordered pseudoparticle Hamiltonian expansion.
This uses the exact ground state of the problem
as a reference state \cite{Carmelo93}. A central
point is that in the pseudoparticle basis this is a
non-interacting ground state. The above expansion
is in the scattering order of the pseudoparticle
processes and corresponds to a pseudoparticle perturbation
theory. The perturbative character of the pseudoparticle
basis assures its convergence. At high energies
the physics is determined by pseudoparticle
forward-scattering interactions of all orders.
On the other hand, while in the case of small-momentum and
low-energy excitations only the non-interacting
Hamiltonian term $(45)$ is relevant, in the
case of low-energy excitations changing the values
of conserving numbers $N_{\iota}$ both that term and
the two-pseudoparticle interaction term $(49)$ contribute.
The two-pseudoparticle interactions fully control the low-energy
physics, as in a Fermi liquid.

Therefore, the use of the same kind of approaches as
in the theory of Fermi liquids leads to the correct
description of the static and dynamic properties of the system
in a straightforward manner. We have also
studied the conformal spectrum which is nothing
but the low-energy spectrum of the above
normal-ordered pseudoparticle Hamiltonian. The
pseudoparticle basis allows an operator description
for the generators of the Virasoro algebra.

The properties of the present bosonic model are a simple
and clear manifestation of the Landau-liquid character
of its fermionic elementary excitations. However, our
study reveals that one boson of vanishing
energy and momentum decays in {\it all} the pseudoparticles
of the pseudo-Fermi sea. This is a non-Fermi liquid
property which justifies the non-perturbative character
of the bosonic basis.

%%%%%%%%%%%%%%%%%%%%%%%%%%%%%%%%%%%%%%%%%%%%%%%%%%%%%%%%%%%%%%%%%%%
\nonum
\section{ACKNOWLEDGMENTS}

This work was supported principally by the University of Illinois
and C.S.I.C. (Spain). We thank P. Horsch for providing a computer
program and for useful comments, D. K. Campbell, E. H. Fradkin,
V. E. Korepin, and A. A. Ovchinnikov for illuminating discussions,
and the hospitality and support of the U.I.U.C.. Computations were done on
Cray-2 of National Energy Research Supercomputer Center, Lawrence
Livermore National Laboratory and we are greatful for their
support. A. H. C. N. thanks CNPq, Conselho Nacional de Desenvolvimento
Cient\'ifico e Tecnol\'ogico, (Brazil) for financial support.
J. M. P. C. gratefully
acknowledges the hospitality and support of C.S.I.C. (Madrid).
This work was supported in part by NSF Grant DMR91-22385 at the
University of Illinois at Urbana-Champaign.

%****************************************************************
%********************* R E F E R E N C E S **********************
%****************************************************************

$*$ Present address, Department of Physics, TamKang University,
Tamsui, Taiwan 25137.\\

$**$ Permanent address, University of \'Evora, Department of Physics,
Apartado 94, P-7001 \'Evora codex, Portugal.\\

\newpage

\narrowtext
\begin{tabbing}
\sl \hspace{2cm} \= \sl $\xi^1$
\hspace{1.5cm} \= \sl $v$
\hspace{1.5cm} \= \sl $v_0$
\hspace{1.5cm} \= \sl $\chi $\\
$n\rightarrow 0$ \>
$1$ \> $2\pi n$ \> $2\pi n$ \> $1/(2\pi^2 n^3)$\\
$n\rightarrow \infty$ \>
$\pi^{1/2} n^{1/4}$ \> $2n^{1/2}$ \> $2/\pi $ \> $1/(2n^2)$\\
\label{table1}
\end{tabbing}
\vspace{.5cm}      % for preprint version
Table 1 - Limiting values of the parameter $(55)$,
velocities $(58)-(59)$, and compressibility $(65)$ for $c=1$.

%**********************************************************
%************** F I G U R E   C A P T I O N S *************

%**********************************************************

\figure{The pseudoparticle band $\epsilon (q)$ $(46)$
for different values of the density and $c=1$. Note that
the zero-energy level corresponds to the pseudo-Fermi
momenta $q=\pm q_F$ and that the pseudomomemtum axis
extends from $q = -\infty$ to $q = \infty$.
\label{fig1}}

\figure{The $f$ functions (a) $f^1$ and (b) $f^{-1}$
of Eq. $(52)$, and (c) the symmetric and (d)
antisymmetric Landau parameters $(53)$
as a function of density and $c=1$.
\label{fig2}}

\figure{The velocities $v$ $(58)$, $v_0$, and $v_1$ $(59)$ as
a function of the density and $c=1$.
\label{fig3}}

\figure{The dressed charge $\xi^1=1/\xi^0$ $(56)$
as a function of the density and $c=1$.
\label{fig4}}

\figure{The excitation spectrum $(61)$ of the one-pair
pseudoparticle-pseudohole eigenstates for $K>0$,
various values of the density, and $c=1$. For
$K<0$ we note that $E(K)=E(-K)$.
\label{fig5}}

\figure{The inverse of the compressibility $(65)$
as a function of the density and $c=1$.
\label{fig6}}
\end{document}